\documentclass[useAMS,usenatbib,usegraphicx]{mn2e}
\usepackage{graphicx}
\usepackage{color}
\usepackage{amssymb}

\title[Blazhko modulation statistics of RRab stars]{The Konkoly Blazhko Survey: Is light-curve modulation a common property of RRab stars?\thanks{Based on observations collected with the automatic 60 cm telescope of Konkoly Observatory, Budapest, Sv\'abhegy}}
\author[J. Jurcsik et al.]{J. Jurcsik$^{1}$\thanks{E-mail: jurcsik@konkoly.hu},
\'A. S\'odor$^{1}$, B. Szeidl$^{1}$, Zs. Hurta$^{1,2}$, M. V\'aradi$^{1,3}$,  K. Posztob\'anyi$^{4}$ \and K. Vida$^{1,2}$, G. Hajdu$^{2}$, Zs. K\H ov\'ari$^{1}$, I. Nagy$^{2}$, L. Moln\'ar$^{1}$, B. Belucz$^{2}$\\
\\
$^{1}$Konkoly Observatory of the Hungarian Academy of Sciences, H--1525 Budapest PO Box 67, Hungary\\
$^{2}$Dept. of Astronomy, E\"otv\"os University, H--1518 Budapest PO Box 49, Hungary\\
$^{3}$Observatoire de Gen\'eve, Universite de Gen\`eve, CH--1290, Sauverny, Switzerland\\
$^{4}$AEKI, KFKI Atomic Energy Research Institute, Thermohydraulic Department, H--1525 Budapest 114, PO Box 49, Hungary\\
}
\begin{document}

\date{Accepted 2009 ..... Received 2009 ...; in original form 2009 Jul 10}

\pagerange{\pageref{firstpage}--\pageref{lastpage}} \pubyear{2008}

\maketitle

\label{firstpage}

\begin{abstract}
A systematic survey to establish the true incidence rate of the Blazhko modulation among short-period, fundamental-mode, Galactic field RR Lyrae stars has been accomplished. The Konkoly Blazhko Survey (KBS) was initiated in 2004. Since then more than 750 nights of observation have been devoted to this project. A sample of 30 RRab stars was extensively observed, and light-curve modulation was detected in 14 cases. The $47\%$ occurrence rate  of the modulation is much larger than any previous estimate. The significant increase of the detected incidence rate is mostly due to the discovery of small-amplitude modulation. Half of the Blazhko variables in our sample show modulation with so small amplitude that definitely have been missed in the previous surveys. We have found that the modulation can be very unstable in some cases, e.g. RY Com showed regular modulation only during one part of the observations while during two seasons it had stable light curve with abrupt, small changes in the pulsation amplitude. This type of light-curve variability is also hard to detect in other Survey's data.  The larger frequency of the light-curve modulation of RRab stars makes it even more important to find the still lacking explanation of the Blazhko phenomenon.

The validity of the [Fe/H]($P,\varphi_{31}$) relation using the mean light curves of Blazhko variables is checked in our sample. We have found that  the formula gives accurate result for small-modulation-amplitude Blazhko stars, and this is also the case for large-modulation-amplitude stars if the light curve has complete phase coverage. However, if the data of large-modulation-amplitude Blazhko stars are not extended enough (e.g. less than 500 data points from less than 15 nights), the formula may give false result due to the distorted shape of the mean light curve used.
\end{abstract}

\begin{keywords}
stars: horizontal branch -- 
stars: variables: other -- 
stars: oscillations (including pulsations) --
methods: data analysis --
techniques: photometric
\end{keywords}

\section{Introduction}

The light-curve modulation of RR Lyrae stars, the so called Blazhko effect, is a hundred-year-old puzzle of stellar pulsation.  The Fourier spectra of the light curves of Blazhko variables are characterized by modulation frequency series appearing in the vicinity of the pulsation frequency ($f_0$) and its harmonic frequencies ($kf_0, k>1$). Based on the location and number of the independent modulation frequency components some attempts have been made to introduce different subtypes of the modulation ($\nu_1$, $\nu_2$, BL1, BL2, BL2x2, etc. \cite{al03,mo03}), these classification schemes, however, do not indeed help to disclose the unknown triggering mechanism of the modulation, which is most probably common in each Blazhko star. Both observational and theoretical efforts are still needed to understand the modulation of RR Lyrae stars. Though the incidence rate of the modulation was estimated to be about 25-30\% based on visual, photographic and photoelectric observations of Galactic field and globular cluster variables \citep{sm81,sm95,sz88}, in a summary paper on RR Lyrae stars \cite{preston}  wrote: {\it"There is no sharp distinction between singly and multiply periodic variables. When observed with sufficient precision or for long enough intervals of time, all RR Lyrae may be or may become multiply periodic"} -- quoting \cite{bdd}.

More recently, the large CCD surveys e.g. MACHO \citep{al03} and OGLE \citep{mo03,mi03,ss03,co06} found a similar or even smaller fraction of variables exhibiting light-curve modulation than determined by photographic and photoelectric observations. The incidence rates of the modulation in the Large Magellanic Cloud (LMC) and in the Galactic bulge were estimated to be 12-15\% and 23-27\%, respectively. These statistics on Blazhko variables may, however, be biased due to the limits of the photometric accuracy and/or the deficiencies in data sampling. Observation in longer wavelength band, where the amplitude of the modalation signal is small, may also reduce the detection rate of the modulation. The incidence rates determined so far are only lower limits for the occurrence of the modulation according to \citet{co06}. The metal deficiency of the variables was suggested to explain the small fraction of Blazhko stars in the LMC, but \cite{smolec} could not detect any significant difference between the metallicities of RR Lyrae stars in the LMC and in the Galactic bulge.

Hipparcos \citep{es97}, NSVS \citep{wo04}, and ASAS \citep{po05} data bases also provide a large number of RR Lyrae observations but these data are not suitable to estimate the true percentage of variables showing light-curve modulation. Based on the NSVS and ASAS data, \cite{patrick} and \cite{szcz} found modulations in an unrealistically small fraction, about 5\% of RRab stars, RR Lyrae variables pulsating in the fundamental mode, also designated as RR0 stars. Many of the well-known Blazhko variables have not even been detected to show light-curve modulation in the Hipparcos, NSVS and ASAS data.

No systematic survey that aimed to determine the frequency of light-curve modulation of RRab stars in the Galactic field has been performed previously. These variables are bright enough to obtain accurate light curves even with small telescopes. Earlier photometric programs concerning field RRab stars focused on either the light-curve variations of known Blazhko variables or the multicolour light curves of large samples of variables \cite[e.g.][]{sturch,fitch,bo77,lub,sr93} in order to determine their physical parameters (temperature, metallicity, reddening, etc.). Though these latter studies covered a large fraction of the brighter Galactic field variables, the time span, data number, phase coverage, and photometric accuracy of the measurements were not suitable to detect long-period and/or small-amplitude light-curve variations.

The knowledge of the true incidence rate of Blazhko stars has a great importance in finding the correct explanation of the phenomenon, as it shows how common the circumstances that favour the appearance of the modulation are. Exploiting the potential of our full access to an automated photometric telescope, we initiated a systematic study of RRab stars to get an estimate of the incidence rate of the modulation. Though our survey concerns a much more limited sample of variables than were measured in the MACHO and OGLE projects, the photometric accuracy and the denseness of our data sampling allow us to detect the occurrence of light-curve modulation more reliably than in those surveys.  The main results concerning the incidence rate of the modulation of the Konkoly Blazhko Survey are summarized in the present paper.

Results for variables with unstable light curves observed in the KBS that do not have data suitable to discuss in separate papers (RY Com, BD Her and FK Vul) are also published here.

Comparing the photometric and spectroscopic metallicities of Blazhko RRab stars of the KBS sample we also draw some conclusions on the correctness of the photometric metallicity \citep{ju96} derived from the mean light curves of Blazhko variables.

\section{Observations and target selection}

The observations of the KBS began in 2004 using an automated 60 cm telescope (Budapest, Sv\'abhegy) equipped with a Wright Instruments 750x1100 CCD camera (FoV 17'x24') and standard $BVR_{C}I_{C}$ filters. Data were corrected for atmospheric extinction and were transformed to the standard $BVR_CI_C$ photometric system. We obtained accurate, extended, multicolour light curves of fundamental-mode RR Lyrae stars with pulsation periods shorter than half a day. About 4000 hours of measurements were gathered for 30 variables on $\sim 750$ nights during the past five years. Each variable was observed at least as long as it could be undoubtedly decided  whether the light curve was stable on about 20-100 days time base or it was modulated. The typical accuracy of the observations was about 5-20\,mmag, depending mainly on the brightness of the object and on weather conditions. The apparent maximum $V$ brightness of the observed variables was in the 10.5-13.5\,mag range. The accuracy, extension, and denseness of the data enabled us to detect light-curve modulations with small maximum brightness variations (some hundredths of a magnitude), i.e., with Fourier amplitudes of the modulation frequency components in the mmag regime. Data are utilized twofold; accurate multicolour light curves of monoperiodic RRab stars and detailed analyses of individual Blazhko variables have already been and are going to be published.

When selecting stars to observe, special care was taken to choose variables with good comparison stars (close to the variable in projected distance, brightness, and colour) in order to gain the possible most accurate photometry with our instrument. Magnitude differences of the variable relative to a single comparison star were determined and analyzed for most of the stars. The constancy of the comparison stars were verified by measuring several relatively bright stars in the fields. Details for the constancy of the comparison stars are given in the papers discussing the light curve variability of the individual objects \cite[see e.g., Fig. 7 in][]{ju08c}. The ensemble mean brightnesses of five, three and two neighbouring stars were calculated, and used as comparison stars' magnitudes for UZ UMa, SS Cnc and DM Cyg, respectively. We applied the Image Subtraction Method  \citep[ISIS;][]{alard} for the photometry of CZ Lac in order to eliminate the additional light of its close companion.

We have observed only short-period RRab stars ($P<0.5$ days) to have a sample with somewhat homogeneous physical parameters, as most of the physical properties of RRab stars relate, to some degree, to the period of the pulsation. The confinement of our survey to short-period variables also made the observation effective, it assured to reach a good coverage of the entire possible sample within a few years of observations. No star with declination below $5\degr$ was selected to be observed in order to obtain extended data sets.

The philosophy of the target selection was to choose variables with poor quality photometric observations available and to clear up some questionable cases. We endeavoured to obtain the possible most unaffected sample in order to get an unbiased estimate of the incidence rate of the modulation. None of the observed stars was definitely known to show Blazhko modulation previously. In three cases ({RR~Gem}, {MW~Lyr}, and {DM~Cyg}) light-curve modulation was announced based on photographic or visual observations, but these results were in conflict with further observations and/or the reanalysis of the original data could not confirm the formerly detected modulation \citep{so05,so07b}. Our observations revealed that these stars do indeed show light-curve modulation, but with significantly different modulation period, amplitude, and type (amplitude or phase modulation) than that was found previously.  {RY~Com} was suspected to show Blazhko modulation \citep{ju96}, however, the scatter around minimum brightness shown by the available observations \citep{jo66,bo77} might have arisen from photometric inaccuracy. Four of the targets ({TZ~Aur},  {SS~Cnc},  {RR~Gem}, and  {FH~Vul}) were used by \citet{ju96} to calibrate the metallicity versus light-curve parameters' relation assuming that their light curves were stable. Our accurate measurements show that, in fact, two of these stars, {SS~Cnc} and  {RR~Gem} are Blazhko variables.

The 30 RRab stars for which extended photometric data were obtained in the Konkoly Survey comprise about 60\% and 80\% of the possible targets of the Northern Hemisphere summer and winter seasons' RRab stars matching our selection critetia.

\begin{figure}
\begin{center}
\includegraphics[width=8.8cm]{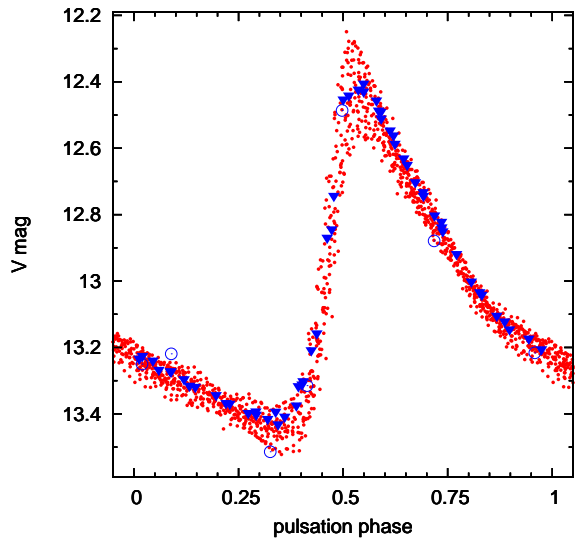}
\end{center}
\caption{Comparison of the Konkoly and previous photoelectric and CCD observations of AQ Lyr. The Konkoly observations are shown by dots, the CCD observations of \citet{cast}, and the photoelectric data of \citet{sturch} are plotted by triangles and circles, respectively. Based on the previous observations no light-curve modulation of AQ Lyr could be detected.}
\label{aqllc} 
\end{figure}

\begin{figure*}
\begin{center} 
\includegraphics[width=14.8cm]{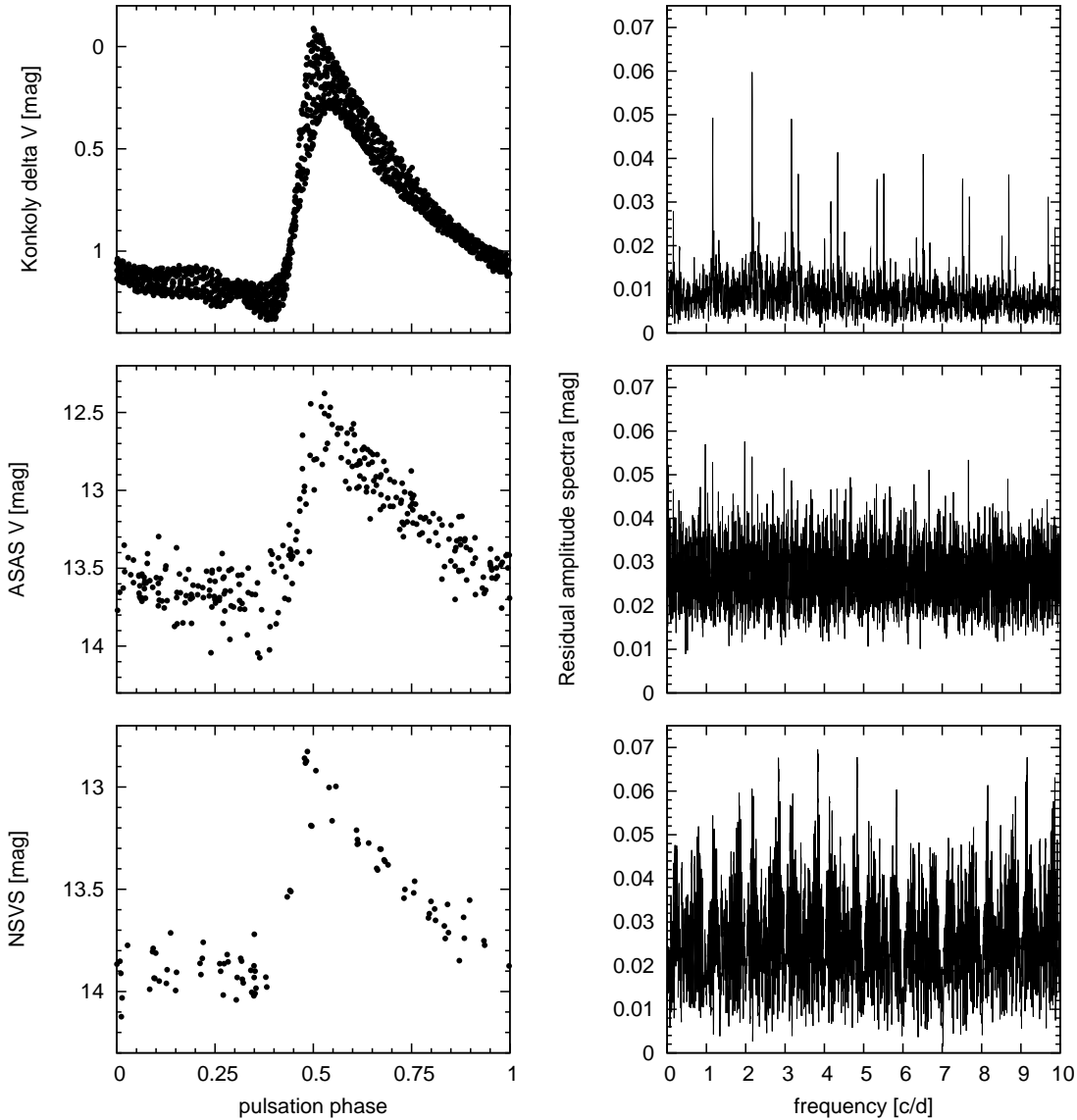}
\end{center}\caption{
Comparison of the Konkoly and previous CCD light curves of UZ Vir. The residual spectra of the data prewhitened for the pulsation light variation are shown in the right-hand panels. The largest amplitude signals in the residual spectrum of the Konkoly data are modulation components at 0.015 cd$^{-1}$ separation from the pulsation frequency components and $\pm 1$ cd$^{-1}$ alias frequencies of these modulation components. Although one of the modulation components ($f_0-f_m$) appears in the residual spectrum of the ASAS data, too, this is not the highest amplitude signal and no other significant peak with the same separation from the pulsation components can be identified. The scarce NSVS data show a scattered light curve.}\label{uzv} 
\end{figure*}

\section{The Konkoly Blazhko Survey}

About half of the 30 observed RRab stars (14 variables, 47\%) were found to exhibit light-curve modulation. This is a much larger percentage than was given by any previous estimate. None of the discovered Blazhko variables could have been undoubtedly identified to show light-curve modulation from the Hipparcos, NSVS, ASAS or any other previous photometry. As examples, our observations are compared to the published light curves for two RR Lyrae stars discovered to show Blazhko modulation in our survey in Figs.~\ref{aqllc} and \ref{uzv}.

AQ Lyrae, one of the stars measured in the survey, was previously supposed to have stable light curve. \cite{cast} observed AQ Lyr on six consecutive nights in 1995 and found no light-curve modulation. The earlier photoelectric observations of \cite{sturch} matched the CCD light curve observed by \cite{cast} within the limits of the photometric accuracy. Our extended data have reveled, however, that in fact AQ Lyr shows large-amplitude modulation. 
Although the variation in its maximum $V$ brightness is $\sim 0.3$ mag, its light curve does not change significantly on a few days time base due to the 65 days length of its Blazhko period. In Fig. \ref{aqllc} our $V$ light curve of AQ Lyr is compared to the previous observations.

In Fig. \ref{uzv} the Konkoly photometry is compared with the ASAS-3 and NSVS light curves of UZ Vir. The Konkoly light curve shows about 0.3 mag variation in maximum brightness, and corresponding to the $P_{Bl}=68$ d period of the modulation many triplet components can be identified with 0.015 cd$^{-1}$ (cycle per day) separation from the pulsation components in the prewhitened spectrum. Despite its strong maximum brightness variation, UZ Vir has not been found to be a Blazhko variable either from the ASAS or the NSVS data. These data show scattered light curves and very high noise level of the residual spectra. Although the $f_0-f_m$ \footnote{The modulation frequency $f_m=1/P_{Bl}$. The primary modulation features in the Fourier spectra of Blazhko stars are triplets at $kf_0-f_m; kf_0; kf_0+f_m$ frequencies.} modulation side-frequency appears in the residual spectrum of the ASAS data, this is not the largest amplitude frequency and there is no other signal in the residual spectrum with the same separation from any of the other pulsation components. 

Table~\ref{rr} summarizes the observations and the results of the 30 RRab stars measured in the Konkoly Blazhko Survey. The identification of the  comparison stars are given in the second column. The third, fourth, and fifth columns list the filters, the number of nights, and the time interval of the observations, respectively. Previous photometric light-curve information on the targets are given in the 'light-curve history' column. Pulsation periods are taken either from the GCVS \citep{gcvs} or from our works (6th, 7th columns in Table~\ref{rr}). The references given in the last column give information on the access to the Konkoly photometric data.\footnote{All the publications of the Konkoly Blazhko group and the Konkoly photometric data can also be downloaded from the website: http://www.konkoly.hu/24/publications .}

\begin{table*}
\begin{minipage}{175mm}
\begin{center}
\caption{Summary of the Konkoly RR Lyrae Survey. Blazhko and monoperiodic variables are typeset in boldface and italics, respectively.}
\label{rr}
\begin{tabular}{lllrrllcc}
\hline
Star
&Comparison Star
&Filters
&\multicolumn{1}{c}{No.\footnote{Number of nights of the observations.}}
&\multicolumn{1}{c}{time\footnote{Time interval of the observations.}}
&light-curve history\footnote{n/a: not enough photometric data. } 
&$ P_0 $ [d]
&\multicolumn{2}{c}{Ref.\footnote{References: 1) forthcoming papers of the Konkoly RR Lyrae group; 2) \cite{ju06b}; 3) \cite{so07a}; 4) \cite{gcvs}; 5:) \cite{ju06a}; 6) \cite{ju05a}; 7) \cite{ju08c}; 8) \cite{so06};  9) \cite{ju08a}; 10) \cite{kun};  11) \cite{ju09b}; 12) \cite{ju08b}; 13) this paper.}}\\
&&&&&&&$P_0 $&data\\
\hline
{\bf AQ Lyr}& GSC2.3 N21I000423 &$BV(I)_\mathrm{C}$  &85&460 &stable                &0.357134 & 1 & 1\\
{\bf V759 Cyg}& GSC2.3 N2IA000839 &$V(I)_\mathrm{C}$   &74&152 &n/a                 &0.360014 & 1 & 1 \\
{\bf SS Cnc}&ensemble mean of 3 stars  &$BV(RI)_\mathrm{C}$ &35&79 & stable &0.367337 & 2 & 2\\
{\it EZ Cep}& GSC 4521-00784 & $BV(I)_\mathrm{C}$ &26&202 &n/a                   &0.3790035 & 3 & 3  \\
{\it BK Cas}& GSC 4025-01395 & $V(I)_\mathrm{C}$  &11&337 &n/a                   &0.3902700 & 3 & 3  \\
{\bf BR Tau}&  NA45000305      & $BV(I)_\mathrm{C}$ &115&843 &n/a                   &0.3905928 & 1 & 1 \\
{\it TZ Aur}& BD +41 1609 & $BV(RI)_\mathrm{C}$&13&29 &stable                &0.3916746 & 4 & 5 \\
{\it ET Per}& GSC 3671-01241 & $BV(RI)_\mathrm{C}$&12&23 &n/a                   &0.3940135 & 3 & 3   \\
{\bf RR Gem}& BD +31 1547 & $BV(RI)_\mathrm{C}$&63&111 &stable/contradictory  &0.3972893 & 6 & 6\\
{\bf MW Lyr}& GSC2.3 N0223233663 & $BV(I)_\mathrm{C}$ &177&361 &contradictory        &0.3976742 & 7 & 7 \\
{\it V378 Per}& GSC2.3 NCGO000977 & $V(I)_\mathrm{C}$  &15&96 &n/a                   &0.3987208 & 9 & 9 \\
{\bf XY And}& GSC2.3 NBXO000560 &$V(I)_\mathrm{C}$   &64&467 &n/a                   &0.398725 & 1 & 1 \\
{\it FH Vul}& GSC2.3 N2P8000417 & $BV(I)_\mathrm{C}$ & 8&108 &stable                &0.405413  & 10 & 10\\
{\it CN Lyr}& GSC2.3 N24S000237 & $BV(I)_\mathrm{C}$ & 8&59 &stable                &0.4113823 & 4 & 10 \\
{\bf DM Cyg}&GSC2.3 N0330220980; N03302207371  &$BV(I)_\mathrm{C}$  &81&446 &contradictory         &0.419863  & 11 & 11 \\
{\it BK And}& GSC2.3 N078000076 &$BV(I)_\mathrm{C}$  &20&99 &n/a                   &0.4216093 & 9 & 9 \\
{\bf CZ Lac}& Image Subtraction Method        &$BV(RI)_\mathrm{C}$ &116&465 &n/a                  &0.432174  & 1 & 1 \\
{\it GI Gem}& GSC2.3 N8N9000652 &$BV(I)_\mathrm{C}$  &22&92 &n/a                   &0.4332664 & 12 & 12  \\
{\bf FK Vul}& GSC2.3 N2P8000417 &$BV(I)_\mathrm{C}$  &14&83 &n/a                   &0.4340529 & 4 & 13 \\
{\it SW CVn}& BD +37 2310 &$BV(I)_\mathrm{C}$  &10&58 &n/a                   &0.441671  & 12 & 12  \\
{\it BH Aur}& GSC 02397-00378 &$V(RI)_\mathrm{C}$  &12&19 &n/a                   &0.4560898 & 4 & 5  \\
{\it UU Boo}& GSC2.3 N6AZ000508 &$BV(RI)_\mathrm{C}$ &16&396 &n/a                   &0.4569339 & 9 & 9   \\
{\bf UZ Vir}& GSC2.3 N5IZ000213 &$BV(I)_\mathrm{C}$  &70&474 &n/a                   &0.4593925 & 1 & 1  \\
{\it CG Peg}& GSC2.3 N2MC000574 &$BV(I)_\mathrm{C}$  &11&95 &stable                &0.4671382 & 4 & 10 \\
{\bf UZ UMa}& ensemble mean of 5 stars &$V$                 &30&115 &n/a                   &0.4668413 & 8 & 8  \\
{\bf RY Com}& GSC2.3 N5CI000130 &$BV(I)_\mathrm{C}$  &98&768 &n/a\footnote{\citet{ju96} noted possible Blazhko modulation. The previous observations  \citep{jo66,bo77} show, however, large scatter at minimum light that may be attributed to photometric inaccuracy. See also Fig.~\ref{ryclc}.}&0.468951 & 13 & 13\\
{\it SU Leo}& GSC2.3 N6WV000233 &$V(I)_\mathrm{C}$   &12&123 &n/a                   &0.4722633 & 12 & 12   \\
{\bf BD Her}& GSC2.3 N2BS000412 &$BV(I)_\mathrm{C}$  &16&69 &n/a                   &0.4739064 & 4 & 13\\
{\it RZ Cam}& GSC2.3 N7T2000280 &$BV(I)_\mathrm{C}$  &17&75 &n/a                   &0.4804514 & 12 & 12   \\
{\it TW Lyn}& GSC 02971-00853 &$BV(RI)_\mathrm{C}$ &17&26 &n/a                   &0.4818600 & 4 & 5 \\
\hline

\end{tabular}
\end{center}
\end{minipage}
\end{table*}

\begin{table}
\begin{minipage}{80mm}
\begin{center}
\caption{Summary of the modulation properties of the 14 Blazhko stars discovered in the Konkoly Blazhko Survey.}
\label{bl}
\begin{tabular}{lcccc}
\hline
Star
&$ P_\mathrm{Bl}$ [d]
&&$ A_\mathrm{Bl}$ [mag] \footnote{$A(V)_\mathrm{max}$: Full amplitude of the maximum light variation in $V$ band; $A(V)$: Fourier amplitude of the largest amplitude modulation frequency component in $V$ band; $A(I_C)$: Fourier amplitude of the largest amplitude modulation frequency component in the vicinity of $f_0$ in $I_\mathrm{C}$ band.}& \\
&&$A(V)_{max}$&$A(V)$&$A(I_C)$\\
\hline
{\bf AQ Lyr}& 64.9\footnote{complex multiperiodicity of the modulation is detected.} & 0.30 & 0.032 & 0.022 \\
{\bf V759 Cyg}& 16.0{\textsuperscript{\small $b$}} & 0.12 & 0.014 & 0.009 \\
{\bf SS Cnc}& 5.3 & 0.09 & 0.015 & 0.009 \\
{\bf BR Tau}& 19.3 & 0.13 & 0.012 & 0.007 \\
{\bf RR Gem}& 7.2 & 0.09 & 0.007 & 0.004 \\
{\bf MW Lyr}& 16.5 & 0.45 & 0.090 & 0.056 \\
{\bf XY And}& 41.4 & 0.20 & 0.050 & 0.031 \\
{\bf DM Cyg}& 10.6 & 0.07 & 0.010 & 0.006 \\
{\bf CZ Lac}& 14.6/18.6 & 0.46 & 0.027 & 0.016 \\
{\bf FK Vul}& $\sim$56 & 0.3 & 0.06 & 0.04 \\
{\bf UZ Vir}& 68.2 & 0.36 & 0.053 & 0.036 \\
{\bf UZ UMa}& 26.7/143: & 0.14 & 0.020 & -- \\
{\bf RY Com}\footnote{modulation properties in 2007.}& 32: & 0.06 & 0.008 & 0.005 \\
{\bf BD Her}& $\sim$22 & 0.5 & 0.12 & 0.08 \\
\hline
\end{tabular}
\end{center}
\end{minipage}
\end{table}
The basic parameters of the modulation of the 14 Blazhko stars discovered in the Konkoly Blazhko Survey are listed in Table~\ref{bl}. The modulation periods and amplitudes were determined from our observations. The modulation amplitudes listed in Table~\ref{bl} give different measures of the strength of the modulation. The full amplitude of the maximum-light variation ($A(V)_{max}$) and the Fourier amplitude of the largest modulation peak in $V$ band ($A(V)$) are listed. To be comparable with the OGLE-II results, the last column gives the $I_\mathrm{C}$ amplitude of the highest modulation peak in the vicinity of the main pulsation frequency ($A(I_C)$).  This frequency is not always identical with the largest-amplitude  modulation-frequency component, as our sample shows that the highest-amplitude modulation component appears in the vicinity of the first or second harmonic components of the pulsation ($2f_0$, and $3f_0$) in about the half of the Blazhko variables.

The observations of the 16 monoperiodic RRab stars of the sample were published by \citet{ju06a,ju08a,ju08b,kun}, and \citet{so07a}. These stars were observed on 8-26 nights, spanning 19-443 days time intervals. Within the limits of our observational accuracy the light curves of these stars were stable. However, we cannot exclude the possibility that these stars may also show short- or long-period "micro-modulation" with 1-2 mmag amplitude of the modulation side frequencies, which is below our detectability limit. According to the compilation of Blazhko stars by \cite{sm95} about half of the short-period Blazhko stars have modulation periods longer than 50 days. Therefore, it cannot be excluded either that some of the stars claimed here to be monoperiodic may have long modulation periods, longer than 3-4 times the time interval of the observations. Moreover, there is a third possibility for the non-detection of the modulation of these stars, namely the temporal behaviour of the modulation as shown for RR~Gem by \cite{so07b} and for RY~Com in the next section of this paper. These stars can be regarded as monoperiodic only with these restrictions in mind.

Detailed analyses of RR~Gem, SS~Cnc, and DM~Cyg, three Blazhko variables showing small-amplitude light-curve modulation that was not detected previously were published by \citet{ju05a,ju06b,ju09b}. The doubly periodic modulation of UZ~UMa was shown by \citet{so06}, while the first really detailed analysis of the light curve and colour behaviour of a large-modulation-amplitude Blazhko variable, MW~Lyrae, was presented by \citet{ju08c,ju09a}. The observations of AQ~Lyr, V759~Cyg, BR~Tau, XY~And, CZ~Lac and UZ~Vir are extended enough to perform detailed analyses of their Blazhko behaviour, which will be the subjects of further separate papers. Results for the remaining three modulated stars, RY~Com, FK~Vul, and BD~Her are presented in this paper.

Only the half of the new Blazhko variables ({XY~And}, {BD~Her}, {CZ~Lac}, {AQ~Lyr}, {MW~Lyr}, {UZ~Vir}, and {FK~Vul}) have modulation amplitudes large enough to be detectable with less accurate and/or less extended observations. Nevertheless, even though the variation in maximum $V$ brightness is larger than 0.15 mag and the largest-amplitude modulation-frequency component has an amplitude larger than 0.02 mag for these stars, they were not detected to show Blazhko modulation in the Hipparcos, ASAS and NSVS data.

Based on the maximum timings collected in the GEOS data base\footnote{http://rr-lyr.ast.obs-mip.fr/dbrr/dbrr-V1.0\_0.php} and our observations, we have found that there is no indication for any strong, irregular period change in any of the mono-periodic RRab stars in the sample. UU~Boo, SW~CVn and probably EZ~Cep show steady period increase, while the period of FH~Vul is probably decreasing. The periods of the other 12 mono-periodic RRab stars are stable with very small if any changes detected. On the contrary, strong, irregular changes are evident in some of the Blazhko stars (e.g, AQ~Lyr, RR~Gem, BD~Her, V759~Cyg). We thus conclude that the pulsation periods of Blazhko variables tend to be less stable than the periods of unmodulated RRab stars. A similar result was derived from the study of the period changes of RR Lyrae stars in M3 by \cite{m3}.

\subsection {The special case of RY Com}
\begin{figure*}
\begin{center} 
\includegraphics[width=18.0cm]{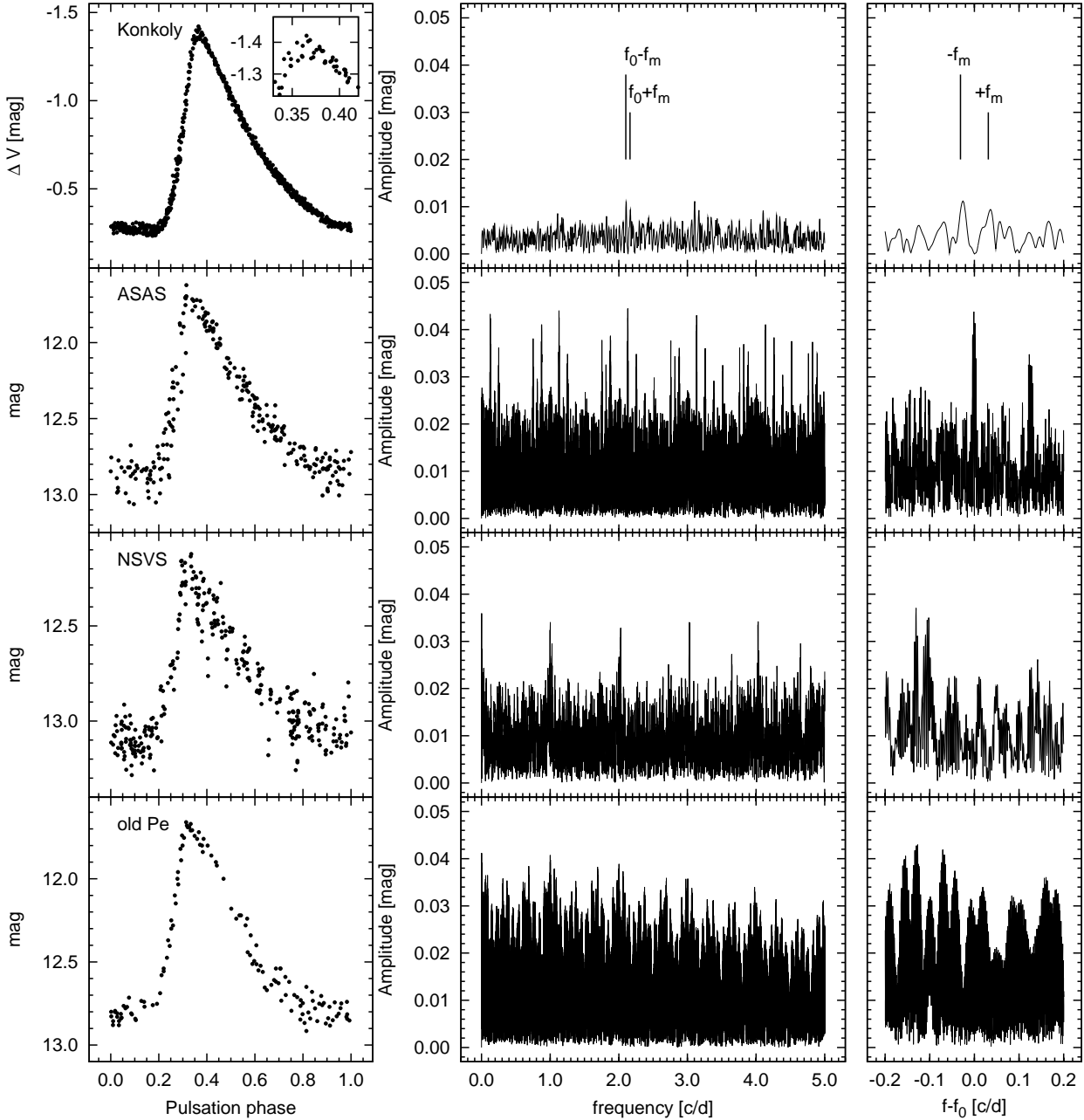}
\end{center}\caption{
Comparison of the Konkoly 2007 light curve and previous photoelectric and CCD observations of RY~Com (right-hand panels). The residual spectra of the data prewhitened for the pulsation period in the $0-5$\,cd$^{-1}$ frequency range and in the $\pm0.2$\,cd$^{-1}$ vicinity of the main pulsation frequency are shown in the middle and right-hand panels, respectively. The residual spectra of the Konkoly observations show small-amplitude ($<0.01$\,mag) modulation-frequency components corresponding to a modulation period of 32\,days. The residual spectra of the other data are dominated by noise and large-amplitude signals due to long-term trends in the data and to pulsation period changes of the star.}\label{ryclc} 
\end{figure*}
\begin{figure*}
\begin{center} 
\includegraphics[width=16.5cm]{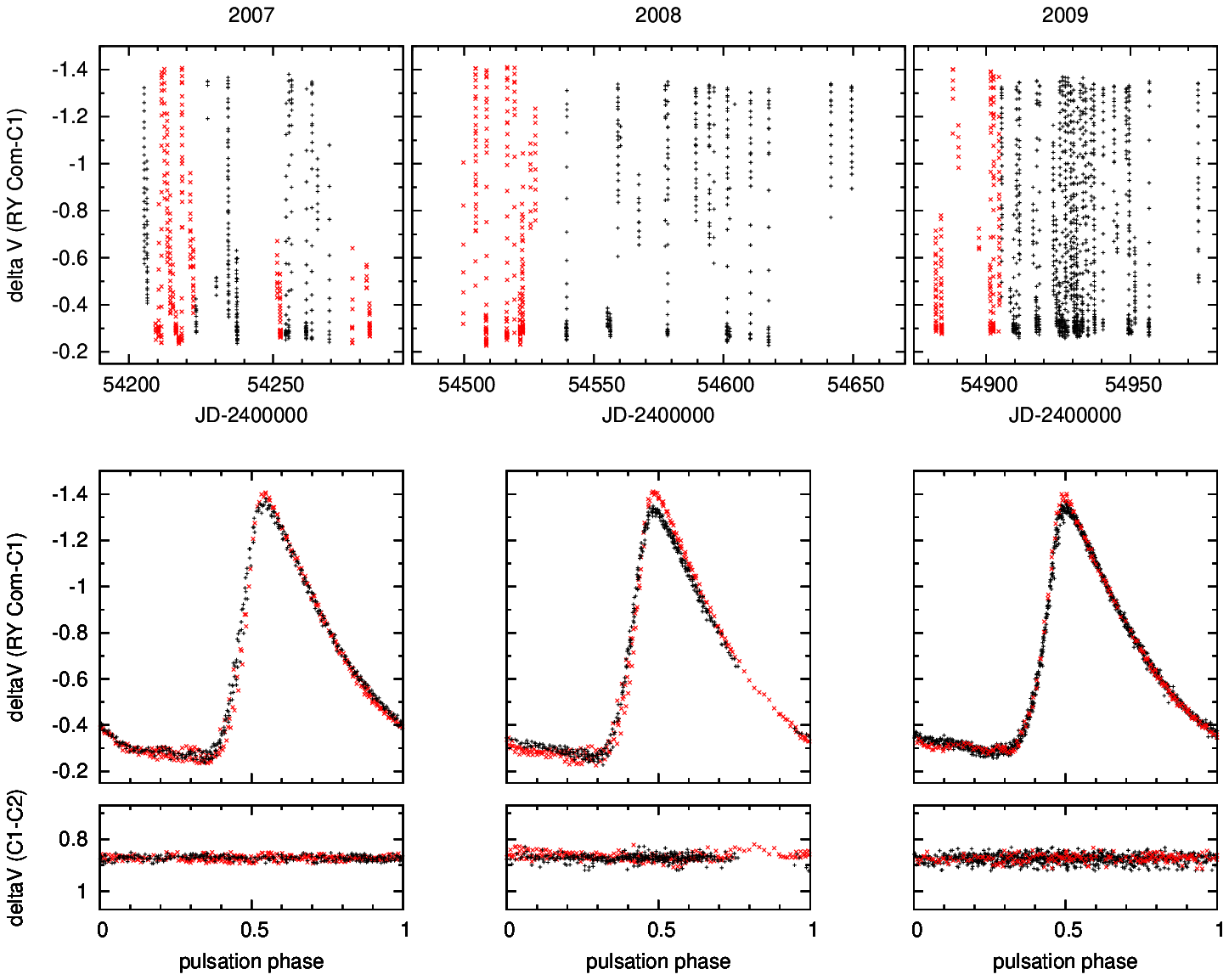}
\end{center}\caption{The Konkoly $V$ light curves of RY Com in the 2007, 2008 and 2009 observing seasons are shown. Crosses denote data belonging to the larger pulsation-amplitude phase of the 32 d modulation detected in the 2007 season and data of the first parts of the 2008 and 2009 observations. Plus signs denote data when the maximum brightness of the star was fainter. The bottom panels show the comparison $-$ check stars magnitude differences (GSC N5CI000130 $-$ GSC3.2 N5CI000137). These data are also phased with the pulsation period of RY Com to demonstrate that the maximum-brightness variation of RY Com does not arise from any defect of the comparison star's magnitudes.  }\label{ryc3} 
\end{figure*}
\begin{figure*}
\begin{center} 
\includegraphics[width=13.6cm]{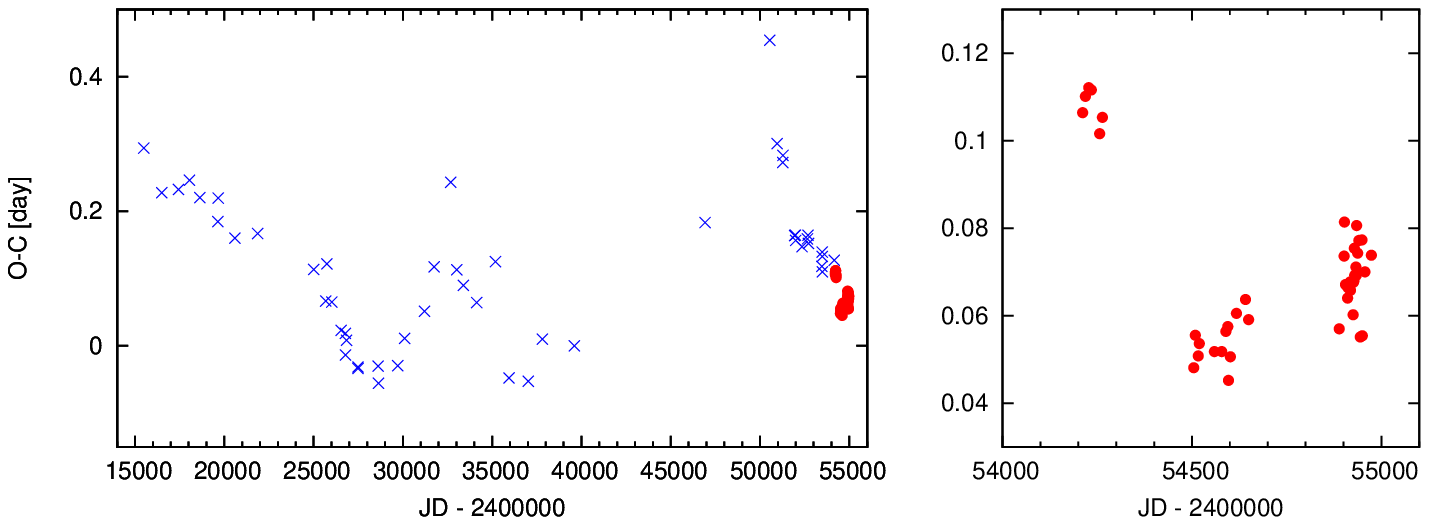}
\end{center}\caption{$O-C$ plot of the maximum timing data of RY Com. Crosses denote data published in the GEOS data base, dots show the recent Konkoly data. The right-hand panel is an enlarged part of the $O-C$ diagram for the last three years using the Konkoly data. The $O-C$ is calculated using 2439598.806 initial epoch and 0.468951 d pulsation period values. }\label{rycoc} 
\end{figure*}

\begin{figure*}
\begin{center}
\includegraphics[width=15.3cm]{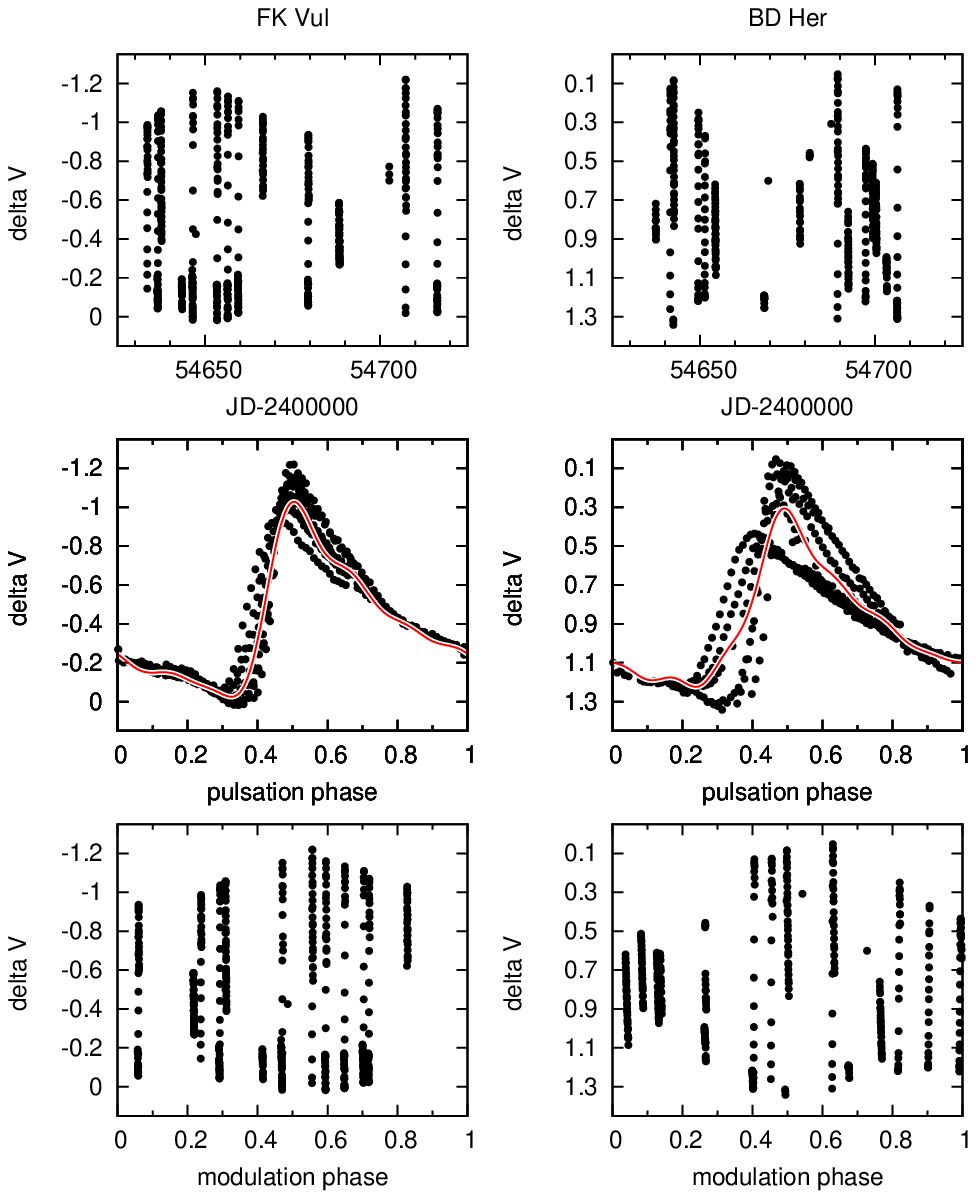}
\end{center}
\caption{Relative $V$ magnitudes of  FK Vul and BD Her versus Julian Date and phased with the pulsation and modulation periods are plotted in the top, middle, and bottom panels, respectively. In the middle panels the mean light curves, determined as appropriate order Fourier fits with the pulsation frequency to the data, are also shown.}
\label{lck}
\end{figure*}

We followed the light-curve variation of RY Com in three consecutive seasons, in 2007, 2008 and 2009.

The Konkoly light curve of RY~Com between JD 2\,454\,205 and 2\,454\,283 (2007 season) is compared with the available photoelectric \citep{jo66,bo77} and CCD (NSVS, ASAS) observations in Fig.~\ref{ryclc}. The scatter of these latter measurements is much larger than the amplitude of the detected light-curve variation in the Konkoly observations. The residual spectra of the four data sets show different characteristics, after prewhitening the data for the mean pulsation light curves. The highest peaks in the residual spectrum of the Konkoly data are modulation frequency components at $f_0\pm f_\mathrm{m}$ ($f_\mathrm{m}=0.03$ cd$^{-1}$) with small, $\sim0.01$\,mag amplitudes, whereas the residual spectra of the ASAS, NSVS and the old photoelectric data show peaks higher than 0.04\,mag at different frequencies. Peaks close to integer frequencies dominate the spectra of the old photoelectric and the NSVS data, indicating long-term trends in the observations, while the residual spectrum of the 4-year long ASAS data shows multiple peaks in the close vicinity of the pulsation frequency components. This is a typical feature if the pulsation period is changing during the time base of the observations. In the ASAS data this is indeed the case, the pulsation period of the variable was 0.00001 d longer during the first 2 years of the ASAS observations than during the second part of the data. Any light-curve modulation of {RY~Com} was definitely hidden in the noise and other biases in all the previous observations.

Hoping to refine the modulation properties of RY Com, we continued its observation in 2008. However, in this season we could not find any sign of a regular modulation, instead, the maximum brightness of the pulsation light curve dropped by about 0.08~mag in a very short period (about a week) at around JD~2\,454\,530. Though the scatter of the phased light curves before and after JD~2\,454\,530 was larger than typical for our observations especially around minimum light, we could not find any definite modulation period in the data. The star behaved similarly in 2009, a sudden jump was detected again in maximum brightness in this season, too, at around JD~2\,454\,905, and no clear modulation was evident. The observed light-curve variations were consistent in the $B, V, I_C$ bands in each season. In Fig.~\ref{ryc3} the light curve history of RY~Com between 2007 and 2009 is documented.

Fig.~\ref{rycoc} shows the $O-C$ variation of RY~Com based on archive data collected in the GEOS data base and on our recent observations. The plots indicate that the pulsation period of RY~Com varies irregularly, showing both abrupt and continuous changes. During the Konkoly observations a period increase of 0.00007 d occurred. Unfortunately, it cannot be resolved from our data whether the period change occurred abruptly at around the drop of maximum brightness (JD~2\,454\,530) or it was continuous.

RY~Com is the second star in our sample with temporarily occurring modulation. For RR~Gem it was found that during the 70 years of the observations the modulation properties of the star changed significantly and that the modulation was not detectable in the 1970-80s \citep{so07b}. The photographic data of RR~Gem showed that the change of the maximum brightness of the mean light curve was most probably connected to an abrupt period change. A 0.00006 d period increase was followed by a 0.1 mag brightening of the $B_{pg}$ mean maximum magnitude of RR~Gem, while the 0.08 mag drop of the mean maximum $V$ brightness of RY~Com was accompanied by a 0.00007 d pulsation period increase in 2008.

\subsection{FK Vul and BD Her}

FK~Vul and BD~Her were observed in the course of the KBS in 2008. Both stars were observed on more than 10 nights. According to these observations large-amplitude modulations of the light curves are evident (see Fig.~\ref{lck}). The observations indicate that the modulation periods of FK~Vul and BD~Her are about 56 and 22 days, respectively. As we do not plan to continue to observe these stars, and the obtained data are not extended enough to allow us to perform detailed analyses of these Blazhko variables, their observations are presented here.

\begin{table}
\begin{center} \caption{Konkoly CCD observations of RY~Com, FK~Vul, and BD~Her. Standard $BVI_C$ magnitude and $B-V$, $V-I_C$ colour differences are given relative to the GSC2.3 stars N5CI000130 (13 05 16.84 +23 16 54.0), N2PT000030 (20~52~40.50 +22~26~25.4), and N2BS000413 (18 50 32.84 +16 27 28.0) for RY~Com, FK~Vul, and BD~Her, respectively. The full table is available as Supporting Information with the online version of this article.}
\label{photdata}
\begin{tabular}{lccc}
\hline
 HJD & delta mag & filter/colour & star \\
$2400000 +$& &&\\
\hline
  54205.28742  & $-1.911$& B & RY\_Com\\
  54205.29319  & $-1.862$& B & RY\_Com\\
  54205.29978  & $-1.835$& B & RY\_Com\\
...&... &...&...\\
\hline
\end{tabular}
\end{center}
\end{table}

\begin{table}
\begin{center}
\caption{Maximum timings derived from the Konkoly CCD $V$ observations for RY~Com, FK~Vul, and BD~Her. The full table is published as Supporting Material. }
\label{maxtime}
\begin{tabular}{ccc}
\hline
maximum time & star \\
\hline
54211.368& RY\_Com\\
54218.404& RY\_Com\\
54227.315& RY\_Com\\
...&...\\
\hline
\end{tabular}
\end{center}
\end{table}

The photometric observations and maximum timings of RY~Com, FK~Vul, and BD~Her are available in the online version of the journal as Supplementary Material (see Tables~\ref{photdata}, and~\ref{maxtime} as samples of the electronic data).

\section{Comparison with the MACHO, OGLE-I, and OGLE-II statistics}

We compare our results with the MACHO LMC \citep{al03}, OGLE-I \citep{mo03}, and OGLE-II \citep{co06} Galactic bulge statistics, as only these publications give enough details on the variables. These surveys are based on the observations of 6158, 150, and 1888 fundamental-mode RR Lyrae stars, among them 944, 51, and 544 have periods shorter than half a day. The detected incidence rates of the modulation of these short-period variables are 16\%, 27\%, and 30\%, while 47\% of the Konkoly sample exhibit the Blazhko effect.

We do not make any distinction between the different types of modulation according to the location of the detected modulation frequency components ($\nu_1$, $\nu_2$, BL1, BL2, BL2x2, etc.), which were used in the other surveys. Our previous results showed that the $\nu_1$, BL1 variables, which are characterized by only one modulation frequency component close to the main pulsation frequency, exhibit, in fact, Blazhko type modulation (BL2, showing equidistant triplet frequencies) with high asymmetry in the amplitudes of the side lobe frequencies \citep{ju05b}. We have also shown that the side lobe frequencies of an amplitude and phase modulated harmonic signal can naturally have highly asymmetric amplitudes \citep{szb}.
\cite{co06} obtained a significantly higher ratio of BL2 stars to BL1 stars than what had been found in the previous surveys. This was attributed mainly to the robustness and sensitivity of the method they applied to detect additional symmetrical frequency components. Their result also indicates that the detection of  equidistant frequency components on both sides of the pulsation frequency depends on the accuracy of the data and the efficiency of the analysis performed. It seems that characterizing the modulation with frequency doublets instead of triplets does not reflect real differences of the modulation. The $\nu_2$, BL2x2 variables show multiperiodic modulation, which is probably also a common feature of Blazhko stars.

\begin{figure}
\begin{center}
\includegraphics[width=8.9cm]{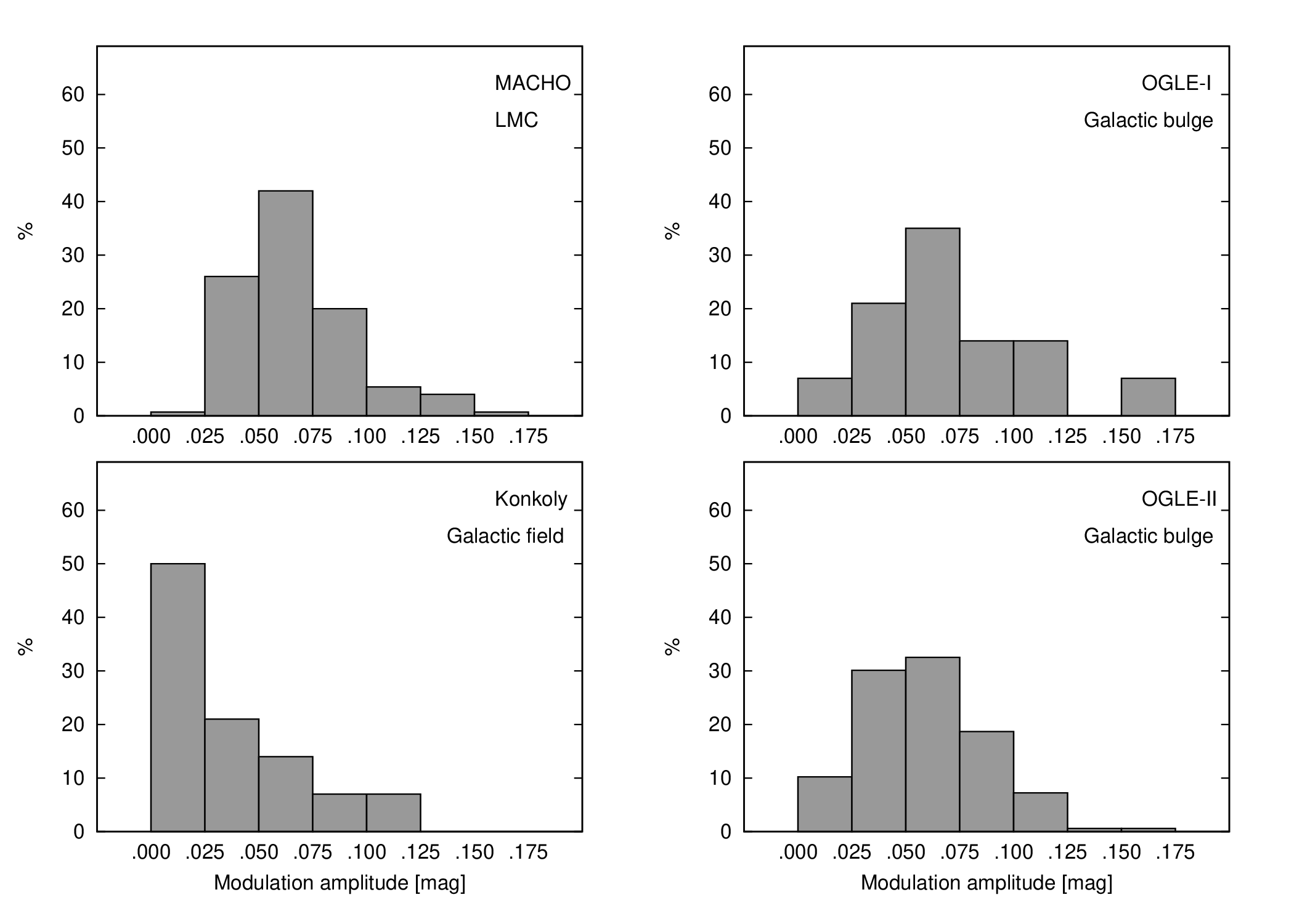}
\end{center}
\caption{Modulation-amplitude distribution of the Blazhko variables with pulsation-periods shorter than 0.5\,days for the MACHO \citep{al03}, OGLE-I \citep{mo03}, OGLE-II \citep{co06}, and Konkoly surveys. There are 148, 14, 166, and 14 short-period Blazhko stars detected in the four surveys. The modulation amplitude corresponds to the $V$ Fourier amplitude of the largest-amplitude modulation-frequency component.  The OGLE  $I_\mathrm{C}$ amplitudes are multiplied by 1.58 in order to be comparable with the $V$ band amplitudes.  The MACHO and OGLE-I,II data show similar distributions, the modulation is the most frequent with 0.050-0.075\,mag amplitudes according to these surveys. On the contrary, the more densely sampled and more accurate Konkoly data show that the modulation has small amplitude ($A_\mathrm{mod}<0.025$\,mag) the most frequently. The Konkoly results warn that a large fraction of Blazhko stars, those with small modulation amplitudes, has  escaped detection previously. 
}
\label{amp}
\end{figure}

The distribution of the modulation amplitude of short-period ($P_\mathrm{puls}<0.5$\,days) Blazhko variables are shown for the MACHO, OGLE-I, OGLE-II, and Konkoly data in Fig.~\ref{amp}. The amplitude corresponds to the largest-amplitude modulation-frequency component in $V$ band. In order to have homogeneous data, the $I_\mathrm{C}$ band amplitudes of the OGLE-I and OGLE-II data are transformed to $V$ amplitudes according to the formula: $A(V)/A(I_\mathrm{C})=1.58$ \citep{ju05c}. Here we stress again that amplitudes of modulation frequency components only in the vicinity of the main pulsation frequency were published for the OGLE-II data, but the largest-amplitude modulation-frequency component occurs in the vicinity of the higher harmonics of the pulsation frequency in some Blazhko variables.

While the MACHO and OGLE data show similar modulation amplitude distributions with the most frequent amplitude of the modulation being between 0.050--0.075 mag, small-amplitude modulations ($A(V) < 0.025$\,mag) are the most frequent in the Konkoly sample. Blazhko variables with such small modulation amplitudes constitute only  1\%, 7\%, and 10\% of the MACHO, OGLE-I, and OGLE-II Blazhko samples, respectively, but 50\% of the modulated RRab stars discovered in the Konkoly Survey belong to this group.

The $V$ and $I_\mathrm{C}$ Fourier amplitudes given in the last two columns in Table~\ref{bl} can be directly compared with the modulation amplitudes measured in the MACHO and OGLE-II surveys. In our sample of fourteen Blazhko stars, six variables have $A(V)\le0.015$ mag and $A(I_\mathrm{C})\le0.009$ mag modulation amplitudes. Modulations with similarly small amplitudes were not detected in the MACHO and OGLE-I surveys at all, and were detected only in 2\% of the short-period RRab stars of the OGLE-II data.

If only Blazhko variables with at least one large-amplitude modulation-frequency component ($A(V)>0.025$\,mag) were considered in the Konkoly data, the incidence rate of the modulation would be 23\%, similarly to the MACHO and OGLE results. Consequently, the differences in the observed incidence rates of the modulation between our and the MACHO and OGLE statistics do not arise from the small number of objects and/or any bias in the target selection of our survey but from the fact, that the latter surveys could not detect small-amplitude light-curve modulations.

As our survey concerns only 30 objects, which is statistically a small number, it is important to verify the significance of the results obtained on this sample. In the OGLE-II data used by \citet{co06} there were 544 RRab stars with $P_\mathrm{puls}<0.5$\,days and among them 30.5\%, 166 variables were detected to show light-curve modulation. Assuming that the true incidence rate of the modulation among RRab stars equals to this ratio, the likelihood of detecting 14 out of 30 variables to show light-curve modulation would be only 4.6\%. Therefore, the small sample size of the KBS cannot be accounted for the large percentage of Blazhko variables detected.

We have also checked whether the selection procedure that \citet{co06} applied to detect Blazhko variables could or could not identify small-amplitude modulations. The last column in Table~\ref{bl} lists the $I_\mathrm{C}$ amplitudes of the highest modulation-peaks in the vicinity of  the main pulsation frequency of the Blazhko variables observed in the Konkoly Survey. To classify a variable to BL1, this peak  has to be as large as 2.4 times the noise level. For a BL2 variable two peaks with symmetrical frequency-separations are required, each with an amplitude larger than 1.1 times the noise level according to the selection applied by \citet{co06}. The average noise level of the light curves of the short-period  Blazhko variables in the OGLE-II sample is 16.7 mmag $\pm$ 8.3\,mmag. According to the data in Table~\ref{bl} the $A(I_\mathrm{C})$ amplitudes of the small-modulation-amplitude variables are about the half of the average noise level of the OGLE-II data.

The probability of detecting  {RR~Gem}-type, small-amplitude modulation in the OGLE-II survey was also tested using different subsets of our observations.  The same number of data points (250) as the OGLE-II light curves contain were randomly selected from the $I_\mathrm{C}$ band light curve of RR Gem. Following the method of \citet{co06} we searched for modulation frequency components in the vicinity of the main pulsation component in the test data sets. As RR Gem has a modulation period of 7.2\,days, modulation frequency components were searched in a wider frequency range ($f_0\pm0.2$\,cd$^{-1}$) than it was done by \citet{co06}.  {RR~Gem} was detected only twice from 1000 random selections to be a Blazhko variable according to the criteria of \citet{co06}. Therefore, we conclude that week modulations like {RR~Gem} has were definitely missed in the OGLE-II survey.

Another reason of the underestimation of the number of Blazhko variables in the OGLE-II data lies in the detection criteria that \citet{co06} applied. They searched only for modulations with $P_\mathrm{mod}>10$\,days period. Taking into account that in our survey 2 of the 14 new Blazhko stars have modulation periods shorter than 10 days, this selection limit also leads to miss some fraction of the Blazhko variables.

Concerning the MACHO data, the standard deviations of the residuals after the removal of the pulsation and modulation components from the light curves are in the 0.03-0.20\,mag range. This very high noise level makes the detection of small-amplitude modulations very unlikely. Using the light-curve solution of RR~Gem and the timings of the MACHO observations we generated artificial time series to test the detection probability of RR~Gem-like modulation in the MACHO data. To mimic the noise properties of the MACHO data, Gaussian noise was added with 0.03, 0.06, and 0.09\,mag rms statistics. Fourier analysis of the test data (prewhitened for the pulsation) showed that  the higher modulation peak in the vicinity of the main pulsation frequency had an amplitude larger than $2\sigma$ of the spectrum only if the smallest (0.03\,mag) rms noise was added. If the rms was 0.06\,mag, there was none among 500 test time series that had a modulation peak reaching $2\sigma$ amplitude. For comparison, only 5\% of the MACHO light curves have rms residual smaller than 0.06\,mag.

Our survey is confined only to short-pulsation-period variables, which tends to bias the sample in favour of metal-rich disk RR Lyraes.
Therefore our conclusion that the true incidence rate of the
modulation is significantly larger than it was estimated in the
previous surveys is valid only for this population of RRab stars.
As the MACHO and OGLE surveys do not show, however, large differences in the occurrence rates of the modulation at different pulsation periods, it is a sound guess that the situation is the same for longer period variables, too. To confirm this supposition accurate photometric data are needed for longer period RRab stars, as well.

\begin{table*}
\begin{center}
\caption{Comparison of spectroscopic and photometric [Fe/H] values of Blazhko stars. Variables with small and large modulation amplitudes are typeset in italics and boldface, respectively.} 
\label{feh}
\begin{tabular}{lrrrcl}
\hline
star& No. of $V$ data
&\multicolumn{2}{c}{$\mathrm {[Fe/H]_{sp}}$ (weight)$^{a}$}&
$\overline{\mathrm {[Fe/H]_{sp}}}$ $^{b}$
&\multicolumn{1}{l}{$\mathrm {[Fe/H]_{phot}}$}\\
&&\multicolumn{1}{r}{\cite{suntzeff}} & \cite{layden}&&\\
\hline
{\it SS Cnc} &       1400    &$-0.16$ (4.0)&$ 0.13$    (2.5)&$-0.03$&$-0.12$\\
{\it RR Gem} &       3000    &$-0.14$ (4.5)&$-0.13$    (2.0)&$-0.14$&$-0.16$\\
{\it DM Cyg} &       3000    &$-0.16$ (1.0)&$ 0.07$    (2.0)&$-0.01$&$-0.01$\\
{\it RY Com} &       1000    &$-0.64$ (0.5)&$-1.38$    (1.0)&$-1.13$&$-1.16; -1.16;-1.08; -1.23; -1.16^{c}$\\
\hline
{\bf AQ Lyr}      &1400   &$-0.19$ (2.0)&             &        &$-0.05$\\
{\bf XY And}      &2600   &             &$-0.68$ (1.0)&        &$-0.68$\\
{\bf CZ Lac}      &8000   &$-0.13$ (2.0)&             &        &$-0.26$\\
{\bf FK Vul}$^{d}$ &500   &             &$-0.71$ (1.0)&        &$-0.37$\\
{\bf BD Her}$^{d}$ &400   &$-0.13$ (2.0)&             &        &$-1.05$\\
 \hline
\multicolumn{6}{l}{\footnotesize $^{a}$ Spectroscopic data transformed to the metallicity scale of the photometric [Fe/H] and their weights.}\\
\multicolumn{6}{l}{\footnotesize $^{b}$ Weighted means as specified in \cite{ju96}.}\\
\multicolumn{6}{l}{\footnotesize $^{c}$ Derived from the mean light curves of the 2007 and 2008, 2009 data before and after JD 2454530 and JD 2454905, respectively.}\\
\multicolumn{6}{l}{\footnotesize $^{d}$ Scarced data.}\\
\end{tabular}
\end{center}
\end{table*}

\section{The photometric metallicity of Blazhko variables}

The metallicity of RRab stars can be derived from the period of the pulsation and the $\varphi_{31}$ epoch independent phase difference of the $V$ light curve with similar accuracy to that determined from low dispersion spectroscopic observations \citep{ju96,kv05}. However, we do not know whether this relation also holds for the mean light curves of Blazhko variables. If the phase coverage of both the pulsation and the modulation cycles is not complete enough then the mean light curves of large-modulation-amplitude Blazhko variables can be seriously distorted, giving rise to spurious value of the Fourier parameters. For instance, using the ASAS data of SS~For \cite{kv05} found that the photometric metallicity deviates significantly from the spectroscopic value, however, the recent, extended photometric data \citep{ssfor} define a mean light curve with Fourier parameters well matching the [Fe/H]($P,\varphi_{31}$) relation.

The extended data we obtained for Blazhko variables enable us to check the accuracy of the photometric metallicities derived from the mean light curves of these variables. Among the 14 new Blazhko variables nine have spectroscopic abundance determinations. Table~\ref{feh} lists these variables and summarizes their spectroscopic and photometric metallicities. In this sample there are four and five stars showing small- and large-amplitude modulations, respectively. The photometric metallicities of the small-modulation-amplitude variables calculated from well defined mean light curves are in very good agreement with the spectroscopic values. We thus conclude that the photometric metallicity of small-modulation-amplitude Blazhko variables is correct, and it has the same accuracy as for unmodulated RRab stars if the light variation is sampled properly.

The situation for Blazhko variables showing large-amplitude modulations is, however, not so obvious. Unfortunately, the spectroscopic information on these stars is less certain than for the small-modulation-amplitude variables. For those variables that have good data coverage (AQ~Lyr, XY~And and CZ~Lac) the spectroscopic and photometric metallicities agree reasonably well. A similar result was derived by \cite{smolec} too, using published photometric and spectroscopic data of five large-modulation-amplitude Blazhko stars. For the two large-modulation-amplitude variables in our sample, which have only 400-600 measurements (FK~Vul and BD~Her), however, the discrepancy between the photometric and spectroscopic metallicities is large. The mean light curves used to calculate the [Fe/H]$_{\rm phot}$ for FK~Vul and BD~Her are also drawn in the middle panels of Fig.~\ref{lck}. It can bee seen that the descending branch of the light curve of FK~Vul is undersampled, which may result in improper shape of the mean light curve. Also, the mean light curve of BD~Her, especially around minimum light, has an unusual shape most probably due to unequal data sampling. It is an open question, however, that if the light curves of FK~Vul and BD~Her were better sampled, then their photometric [Fe/H] would or would not match their spectroscopic values better. The good agreement between the photometric and spectroscopic metallicities of those large-modulation-amplitude Blazhko variables which have good light-curve coverage indicates, however, that if the light variation is well sampled then the Fourier parameters of the mean light curve can be safely used to calculate the metallicity. Nevertheless, to make this statement better grounded, both extended photometric data of other large-modulation-amplitude variables and spectroscopic metallicities of these variables (e.g. for MW~Lyr) are needed.

\section{Summary}
\begin{itemize}

\item
{We have shown that with the inclusion of small-amplitude modulations the incidence rate of the modulation increases to about 50\% for short-period RRab stars. The Konkoly Blazhko Survey showed that, in fact, small-amplitude modulations are equally frequent as large amplitude ones. The distribution of the amplitudes of the modulation show an increase towards smaller amplitudes according to our data. It suggests that the light curves of those RRab stars that are now supposed to be stable may also be modulated but with very small amplitude, below our present detection limit. Moreover, we may have missed detecting small-amplitude modulations with long modulation periods in our sample. Therefore, it is likely that the incidence rate of the modulation we detected is still a lower limit to the true fraction of the RRab stars showing the Blazhko effect. Modulations with amplitudes of the maximum brightness variation in the millimagnitude regime may still have escaped detection. The observations of recent space missions (MOST, CoRoT, Kepler) will clear this question soon. 
}

\item
{There is another embarrassing feature of the Blazhko modulation, namely its temporal occurrence as documented recently in the study of the long term behaviour of  {RR~Geminorum} \citep{so07b} and in the case of RY Com in the present paper. The modulation of  {RR~Lyrae} is also known to cease for a short period in every fourth year \citep{sz88}. Based on these results there is a good chance that the modulation is a common feature of the pulsation of fundamental-mode RR Lyrae stars, it may be even a universal property of these variables. This possibility warns that understanding the Blazhko modulation is crucial in the knowledge of the pulsation of RR Lyrae stars.
}
\item
{Detecting small-amplitude light-curve modulation of RRab stars is also relevant from the theoretical point of view. \cite{kv95} used the lack of small-amplitude modulations as an argument against the oblique magnetic rotator model of the Blazhko phenomenon \citep{sh00}. He says "{\it Assuming random distribution [of the aspect angle], we should see some small- and some large-amplitude modulations. This is apparently not the case}." The discovery of small-amplitude modulation contradicts this argument. Whatever the cause of the Blazhko modulation, it manifests itself in the full possible range of amplitudes. At the same time, simultaneous multi-periodic modulation with similar modulation amplitudes were also detected in the KBS. The significant difference between the modulation periods observed in these stars is a strong argument against those explanations of the Blazhko phenomenon that connect the modulation period to the rotation of the star.
}
\item
{Based on our photometric data and spectroscopic information on Blazhko stars we have found that the photometric metallicity determined from the mean light curves of Blazhko variables are correct if the light curve is properly sampled, i.e., the mean light curve is not distorted by uneven data sampling of the different Blazhko phases. 
}
\end{itemize}

In the Konkoly Blazhko Survey we obtained extended data on 30 RRab stars with fundamental periods shorter than 0.5 days, 14 of them showing the Blazhko effect. This is a significantly larger incidence rate of the modulation than found previously. We cannot exclude the possibility, however, that the large percentage of Blazhko stars in the sample of the KBS is related to the short pulsation period of these stars. If this were the case, the incidence rate of the modulation would 
also depend on the metallicity of the stars as short period RRab stars are, in general, less metal poor than longer period variables.
In order to decide whether there is a period and/or metallicity dependence of the occurrence rate of the modulation we initiated a similar survey focusing on RRab stars of longer periods in 2009.

\section*{Acknowledgments}
We would like to thank Horace Smith for obtaining 6 and 10 nights of observations of MW~Lyr and V759~Cyg with the 60 cm telescope of the MSU.
We are also grateful to Arne Henden for the standard magnitude calibration of the stars in the fields of MW Lyr and CZ Lac.
The constructive comments of the anonymous referee are highly appreciated.
The financial support of OTKA grant K-068626 is acknowledged.
Zs.K. is a grantee of the Bolyai J\'anos Scholarship of the Hungarian Academy of Sciences.

\end{document}